\documentclass{article}
\usepackage{graphicx}
\usepackage{amsmath}

\begin{document}
 
\baselineskip 18pt

\begin{center}
\vspace*{0.5cm}
{\Large {\bf  Neutrino Astrophysics in the cold: Amanda, Baikal and IceCube}}
 \vskip5mm Christian Spiering \footnote{Email address: 
christian.spiering@desy.de}\\
\vspace{1mm} 
DESY, Platanenallee 6, D-15738 Zeuthen, Germany

\bigskip

\begin{abstract}

This talk reviews status and results from the two presently
operating underwater/ice neutrino telescopes, NT-200 in Lake
Baikal and AMANDA-II at the South Pole. It also gives a description
of the design and the expected performance of IceCube,
the next-generation neutrino telescope at South Pole.

\bigskip
\end{abstract}
\end{center}

\section{Introduction}

More than four decades after deep water Cherenkov telescopes for
high energy neutrinos have been proposed \cite{Markov},
two detectors of this type are successfully taking data:
NT-200 in Lake Baikal, and AMANDA-II at the South Pole.
First components of AMANDA's follow-up 
project of cubic-kilometer size, IceCube, 
have been deployed in January 2005.

The science topics of these projects
include the search for steady and variable sources of high
energy neutrinos like Active Galactic Nuclei (AGN), Supernova
Remnants (SNR) or microquasars, as well as the search
for neutrinos from burst-like sources like 
Gamma Ray Bursts (GRB) \cite{Halzen-Stockholm}.
Underwater/ice telescopes
can also be used to tackle a series of questions 
besides high energy neutrino astronomy. 
These include
the search for neutrinos from
the decay of dark matter particles (WIMPs) and the search
for magnetic monopoles or other exotic particles - like strange 
quark matter or Q-balls (see for reviews \cite{WIMP,MM}). 
All these topics are addressed not only by Baikal/NT-200, AMANDA
and IceCube, but also by the Mediterranean projects
reviewed in the talk of J.J.~Aubert at this conference 
\cite{Aubert-Stockholm}.

For deep ice detectors there are two modes of
operation which are not -- or nearly not  --
accessible for detectors in natural water.
Firstly, due to the low light activity of the
surrounding medium, the photomultiplier (PMT) count rate in ice is
only about 
1 kHz. This enables the detection of the feeble increase of 
individual PMT rates 
as caused, for instance, by multiple interactions of Supernova
burst neutrinos over time intervals of a few seconds. 
Amanda-II is actually monitoring about 90\%  of the Galaxy for MeV neutrinos
from Supernova explosions. Secondly, deep ice arrays can be
operated in coincidence with surface air shower arrays
\cite{SPASE}.
This allows one to study questions like the mass composition of 
cosmic rays up to $10^{18}$ eV, to calibrate the neutrino
telescope, and
to use the surface detector as a veto against background events.

This paper is organized as follows: In section 2, the design
and performance of the two running large neutrino telescopes
are sketched,
NT-200 in Lake Baikal and AMANDA-II at South Pole.
Section 3 is a synopsis of the physics results obtained so
far by both detectors (for recent summaries see the talks
at the Neutrino 2004 conference \cite{Djilkib, Amanda-II,
IceCube1}. Section 4 desribes design and expected
performance of IceCube. 

\vspace*{0.5cm}

\section{Design of NT-200 and AMANDA}

Underwater/ice neutrino telescopes consist of a lattice of 
photomultipliers (PMs) housed in transparent glass spheres which
are spaced over a large volume in the Ocean, in lakes
(like the Baikal telescope) or in ice (as AMANDA and IceCube).
The PMs record arrival time and amplitude of the Cherenkov light
emitted by muons or particle cascades.

The Baikal Neutrino Telescope NT-200  is operated in Lake 
Baikal, Siberia,  at a depth of \mbox{1.1 km}.
An umbrella-like frame
carries 8 strings, each with 24 optical modules 
(OMs) arranged in pairs. The pressure glass spheres of the
OMs contain 37-cm diameter PMs. The instrumented
volume forms a cylinder of 70~m height and 42~m
diameter. The angular resolution for through-going muons
is about 4$^{\circ}$.

The present, final configuration of the AMANDA neutrino telescope
is named AMANDA-II. It consists of 677
OMs arranged along 19 vertical strings buried in
the glacial ice at the South Pole, with most of the
OMs at depths between 1500
and 2000 m.  The geometric shape of the
array is a cylinder $\sim$500 m high and $\sim$200 m in diameter.
Each AMANDA OM houses an 8-inch Hamamatsu photomultiplier. 
The present angular resolution for muons tracks is 
about 2.5$^{\circ}$.
Figure 1 sketches the configuration of both detectors.

\begin{figure}[htb]
\begin{center}
\includegraphics[width=8cm]{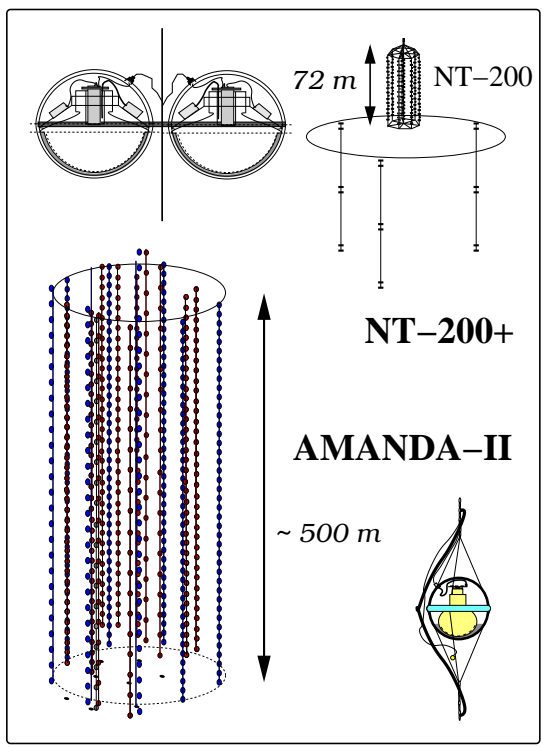}
\label{fig:detectors}
\end{center}
\caption{
AMANDA-II and NT-200 (NT200+). For AMANDA-II, only the central
part between 1500 and 2000 meters is shown, covering about
90\% of the optical modules. Detectors are shown to scale.
Also sketched are the optical modules.
AMANDA PMs have 20\,cm, Baikal PMs 37\,cm diameter. 
}
\end{figure}

\begin{figure}[htb]
\centering
\includegraphics[width=3.2cm]{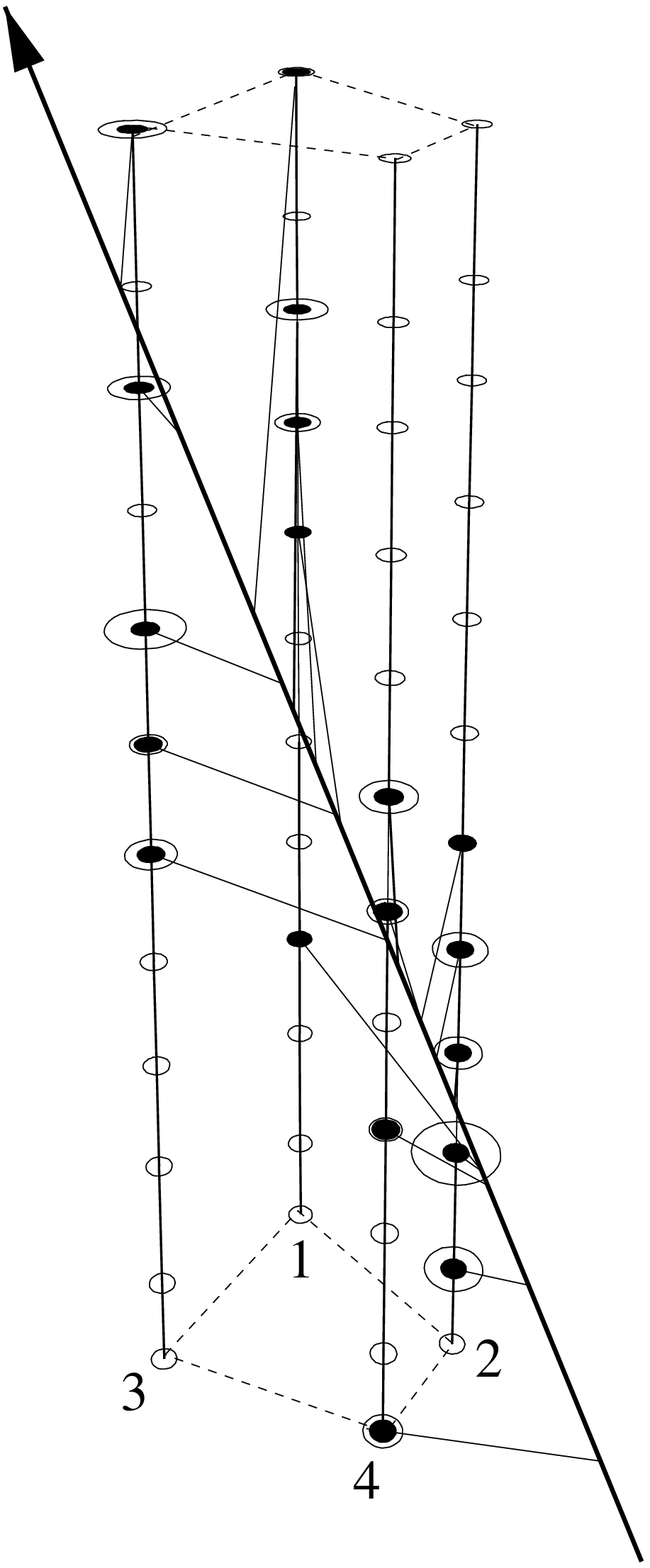}
\hspace{3cm}
\includegraphics[width=3.2cm]{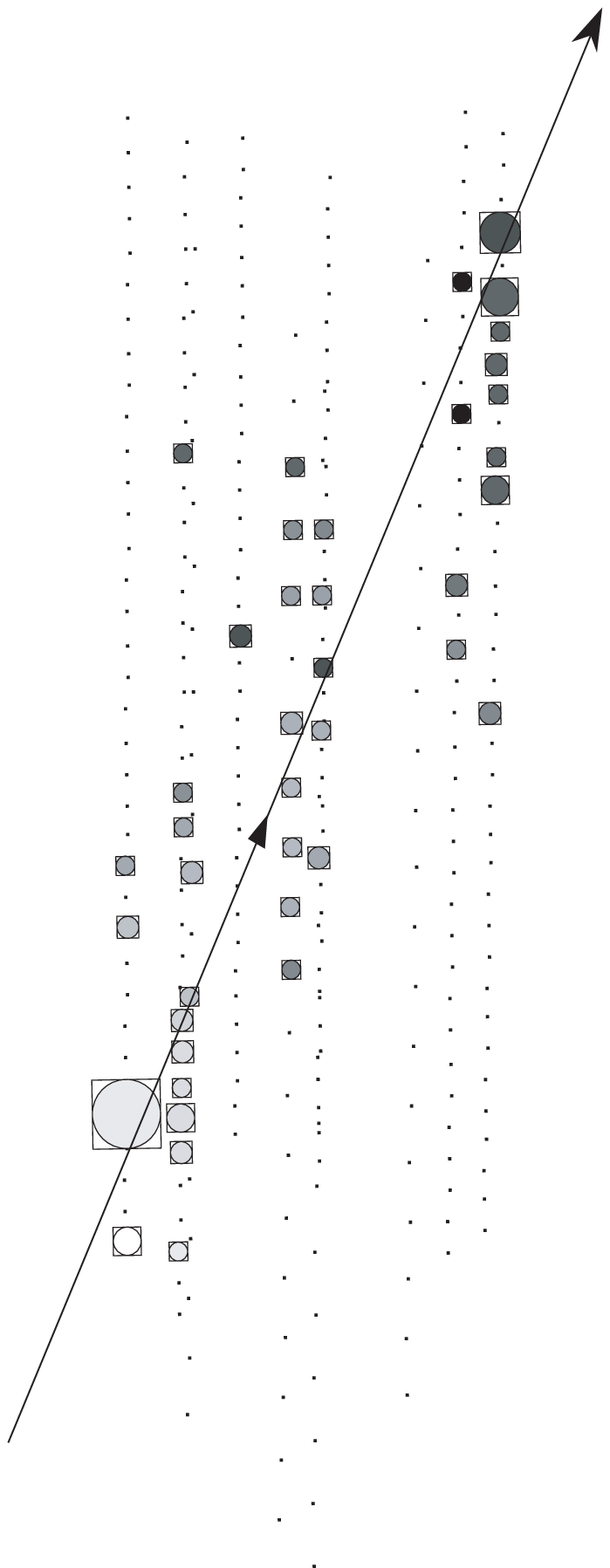}
\caption[1]{
Left: one of the first 
clearly upward moving muons recorded with the 1996 
four-string-stage of the Baikal detector. Black dots denote PMs. 
Hit PMs are encircled, with the size of the disc proportional to the 
recorded amplitude. The arrow line represents the reconstructed muon track,
the thin lines the photon paths.
Right: Upward muon recorded by the 1997 stage of AMANDA. 
Dots denote the PMs arranged at ten strings.  Hit PMs are highlighted,
with the degree of shadowing indicating the arrival time (dark being late),
and the size of the symbols the measured amplitude.
Note the different vertical scales for these events: 70~m
for NT-200, 500~m for AMANDA.
}
\label{fig:nuevents}
\end{figure}

NT-200 takes data since April 1998. It has been
deployed in subsequent stages, starting in 1993
with NT-36, the  pioneering first
stationary underwater array \cite{NT-36}. 
Components are deployed from an natural fixed 
platform: the thick ice layer
which covers Lake Baikal in February and March, when
uutside temperatures reach down to -25$^{\circ}$
(see for more technical details \cite{VLVNT-Baikal}).   
Figure \ref{fig:nuevents}~(left) shows a textbook
neutrino event (an upward moving muon track)
recorded with the early 4-string configuration 
of 1996 \cite{APP2}.

The absorption length of deep Baikal water
varies between 20 and 24 meters.
Since light scattering is small,  NT-200 can monitor a volume
exceeding its own geometric volume by
an order of magnitude. Actually, this is the reason that
the sensitivity of NT-200 with respect to high energy, diffuse-flux
phenomena is not dwarfed by that of the much  larger 
AMANDA-II (AMANDA is embedded in
ice where light scattering is strong and 
diffuses light from distant sources).
In 2005, NT-200 is going to be upgraded with three sparsely
instrumented outer strings (NT-200+, see Fig.1).
The three strings 
will allow a dramatically improved vertex reconstruction
of high energy cascades within this volume
and will increase the sensitivity to diffuse fluxes by a factor
of four (see section 3).

Rather than water, AMANDA uses ice as detection medium. 
The glacial ice is extremely transparent for Cherenkov 
light with wavelengths
near the peak sensitivity of the OMs:
At 400~nm, the average absorption length is 110~m.
Scattering, however, is much stronger than in 
water -- the average effective scattering length is only 20 m.
Below a depth of 1500~m, both scattering and absorption are dominated by
dust, and the optical properties vary with dust concentration.

To deploy AMANDA, holes were melted with hot
water, and strings with OMs frozen into the ice.
Similar to the Baikal telescope, AMANDA was also deployed 
step by step. An earlier 10-string stage,
called AMANDA-B10, was installed
between 1997 and 1999 \cite{NaturePaper}. First neutrinos
have been recorded with the inner four strings deployed in 1996
(see Fig.~2, right). 
AMANDA-II has been taking data since 2000.

\section{Physics results from NT-200 and AMANDA}

\subsection{Proof of principle: atmospheric neutrinos}

\begin{figure}[htb]
\begin{center}
\includegraphics[width=8cm]{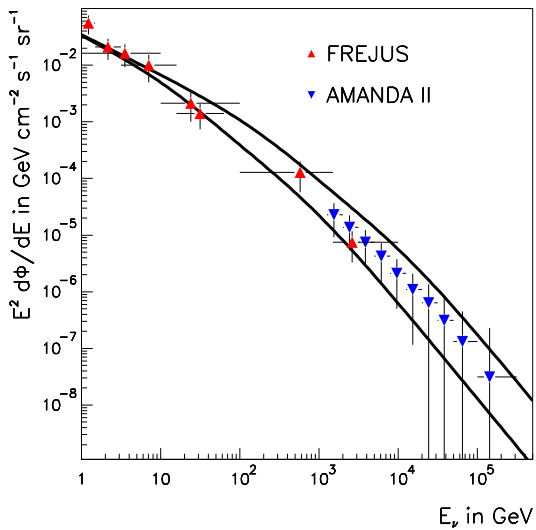}
\vspace{-2mm}
\caption{Atmospheric neutrino energy spectrum (preliminary) from 
regularized unfolding of AMANDA data, compared to the Frejus 
spectrum \cite{Frejus} at lower energies. The two
solid curves indicate model predictions \cite{Volkova} for the horizontal
(upper) and vertical (lower) flux.
}
\label{fig:amaatmu}
\end{center}
\end{figure}

For most analysis channels, both AMANDA and NT-200 use the 
Earth as a filter and separate up-going muons steming from
interactions of neutrinos having crossed the Earth.
The main class of background are
down-going atmospheric muons that are misreconstructed as up-going.
After their rejection, basically up-going muons from
interactions of neutrinos generated in the atmosphere remain.
Atmospheric neutrinos 
not only constitute the main background when  
searching for extraterrestrial neutrinos,
but are also a natural calibration source. 
Figure \ref{fig:amaatmu} shows the  energy spectrum for up-going 
neutrinos based on AMANDA-II data taken in 2000. It has been obtained by
a neural net energy reconstruction,
followed by regularized unfolding. 
This is the first atmospheric neutrino spectrum above a few TeV, 
and it extends up to 300 TeV.

\begin{figure}[htb]
\begin{center}
\includegraphics[width=6.8cm]{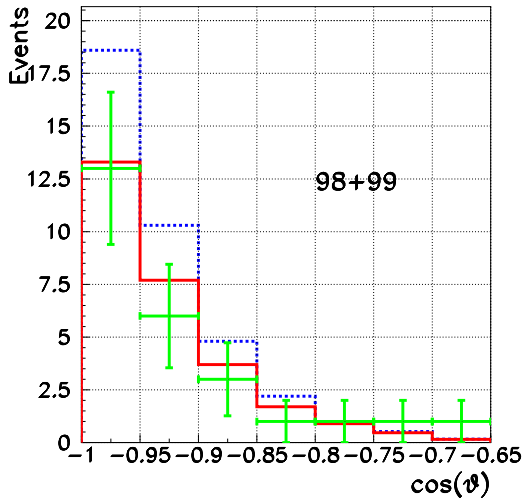}
\vspace{-2mm}
\caption{Angular spectrum of nearly vertical atmospheric neutrinos recorded
with NT-200 in 1998/99. $\cos{\theta} = -1$ corresponds to vertically
upward moving tracks. Data are compared to simulated
distributions including (full line)
and excluding (dotted line) the effect of oscillations.
}
\label{fig:baiatmu}
\end{center}
\end{figure}

AMANDA, with its 60 GeV threshold
for the standard event selection, is not sensitive to neutrino
oscillations. The threshold of the Baikal telescope
NT-200 is much lower and, for tight
selection procedures, can be reduced down to 10 GeV. 
Figure \ref{fig:baiatmu}
shows the angular spectrum for upward tracks close to the
opposite zenit. 
The 20\% deficit close to the vertical is well compatible with the
the oscillation parameters $\delta m^2 \approx 2.5 \cdot 10^{-3}$ and
$\sin ^2 \theta \approx 1$. This effect
has to be taken into account when using
these data as standard signal for detector calibration.

\subsection{Search for a diffuse flux of cosmic neutrinos}

The primary destination of neutrino telescopes is the identification of 
individual, point-like sources of high-energy neutrinos. 
Should individual sources be too weak to produce an unambiguous directional
signal in the array, the integrated neutrino flux from all sources could
still produce a detectable diffuse signal. This flux could
be revealed by an high energy excess on top of the omni-present
background of atmospheric neutrinos.
The data of both experiments have been searched 
for such a diffuse signal using
complementary techniques in different energy regimes.
Event selection is typically optimized
to maximize the sensitivity to an $E^{-2}$ signal spectrum.
Experimental limits given below include statistical as well as
systematic errors (the latter being typically between 20 and
40 percent).

\bigskip

{\bf Upward moving muons:}
The atmospheric neutrino spectrum recorded with AMANDA-II
(Fig.~\ref{fig:amaatmu}) was used to set an upper limit
on a diffuse $E^{-2}$ flux of extraterrestrial muon neutrinos 
for the energy range covered by
the highest bin, 100--300 TeV, by calculating the maximal 
non-atmospheric contribution
to the flux in that bin given its statistical uncertainty.
However, the bins in the unfolded spectrum are correlated and the uncertainty
in the last bin can not a priori be assumed to be Poissonian. The statistics
in the bin was therefore determined with many Monte Carlo samples
used to construct confidence belts. 
Given the (fractional) unfolded number of experimental events 
in the bin,
a preliminary 90\% C.L.\ upper limit of
$E^2 \Phi_{\nu_{\mu}}(E) < 2.6 \times 10^{-7} 
\mbox{cm}^{-2}\mbox{s}^{-1}\mbox{sr}^{-1}\mbox{GeV}$
is derived for $100~\mathrm{TeV} < E_{\nu} < 300~\mathrm{TeV}$.

\bigskip

{\bf Cascades:}
Apart from muon tracks, {\it cascades} can be detected. 
With a typical length of 5-10~m and a diameter of 10~cm, 
cascades can be considered as quasi point-like compared to the
spacing of OMs. All
three neutrino flavors contribute to this signature -- 
cascades stem from
the leptonic vertex of
electron and tau neutrino charged current interactions, and hadronic
vertex cascades from all-flavor neutral current interactions.

The AMANDA analysis focuses on contained events
(allowing good energy reconstruction), 
the Baikal/NT-200 survey to
bright cascades produced at the neutrino interaction
vertex in a large volume around the neutrino telescope.
(As pointed out in the introduction, lack of significant 
light scattering allows to monitor a 
volume exceeding the geometrical volume of NT-200
by an order of magnitude.)

Both analyses did not yield any excess of candidate events
over background: 1 observed event (vs. 0.9 background events)
in the case of the year-2000 AMANDA data, no event 
(vs. 0.4 background events) in the
case of the 1998-2003 NT-200 data. 
The corresponding
upper limits with respect to the flux of all three flavors
are 
$\Phi_{\nu_e +  \nu_{\mu} + \nu_{\tau} }E^2=8.6\cdot10^{-7} 
\mbox{cm}^{-2}\mbox{s}^{-1}\mbox{sr}^{-1}\mbox{GeV}$
(AMANDA, 
$50~\mathrm{TeV} < E_{\nu} < 5~\mathrm{PeV}$ \cite{casc})
and
$\Phi_{\nu_e +  \nu_{\mu} + \nu_{\tau} }E^2=8.1 \cdot10^{-7} 
\mbox{cm}^{-2}\mbox{s}^{-1}\mbox{sr}^{-1}\mbox{GeV}$
(NT-200, 
$22~\mathrm{TeV} < E_{\nu} < 50~\mathrm{PeV}$ \cite{Bai-HE}),
calculated under the assumption that neutrinos arrive
with a ratio
$\nu_e:\nu_{\mu}:\nu_{\tau}=1:1:1$.

\bigskip

{\bf Ultra High Energy neutrinos:}
At ultra-high energies (UHE), above 1 PeV, the Earth is opaque
to electron- and muon-neutrinos. 
The AMANDA search for extraterrestrial UHE neutrinos is therefore concentrated
on events close to the horizon and even from above.
The latter is possible since the atmospheric muon background is low
at these high energies due to 
the steeply falling energy spectrum.
The search for UHE events in 1997 AMANDA-B10 data 
relies on parameters that are sensitive to 
the expected characteristics of an UHE signal:
bright events, long tracks (for muons), 
low fraction of single photoelectron hits.
No excess above background 
is observed and 
a 90\% C.L.\ limit on an $E^{-2}$ flux of neutrinos of all flavors
is derived,
assuming a 1:1:1 flavor ratio at Earth \cite{UHE97}: 
$\Phi_{\nu_e +  \nu_{\mu} + \nu_{\tau} }E^2
< 9.9 \times 10^{-7} 
\mbox{cm}^{-2}\mbox{s}^{-1}\mbox{sr}^{-1}\mbox{GeV}$
($1~\mathrm{PeV} < E_{\nu} < 3~\mathrm{EeV}$).

\vspace{5mm}

\begin{figure}[htb]
\begin{center}
\includegraphics[width=9.5cm,angle=-90]{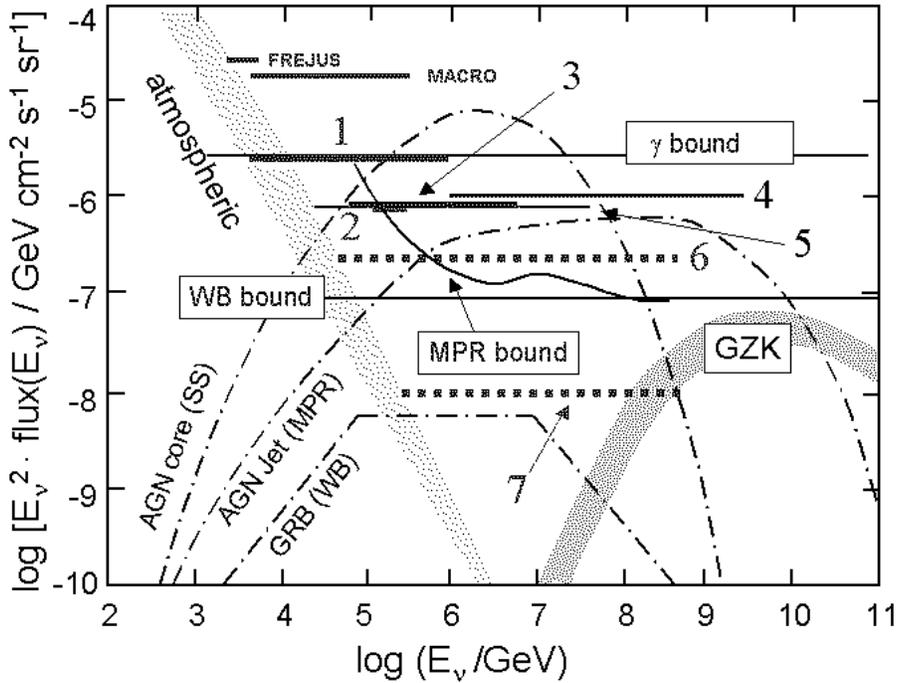}
\caption{
Theoretical and experimental limits and model predictions to
the diffuse flux of extraterrestrial neutrinos, compared to the
flux of atmospheric neutrinos. 
{\it 1)} AMANDA-B10, upward muons \cite{Diff97},
{\it 2)} AMANDA-II, upward muons (prelim.),
{\it 3)} AMANDA-II, cascades \cite{casc},
{\it 4)} AMANDA-B10, UHE events \cite{UHE97},
{\it 5)} NT-200, cascades (prelim.\cite{Bai-HE})
{\it 6)} AMANDA-II and NT200+: 
expectation for 4 years, 
{\it 7)} IceCube expectation for 3 years.
Shown are the limits to the flux of {\it all} flavors,
with the assumption that $\nu_e:\nu_{\mu}:\nu_{\tau}=1:1:1$
at Earth. Experimental limits obtained
for muons alone have been multiplied by a factor of 3.
Theoretical bounds and predictions which have been originally
given for muons and assuming $\nu_e:\nu_{\mu}:\nu_{\tau}=1:2:0$ 
(i.e.without
consideration of oscillations) have been multiplied by
a factor 1.5. For the MPR and WB bound see \cite{MPR,WB}.
}
\label{fig:diffuse}
\end{center}
\end{figure}

{\bf Summary of diffuse searches:}
Using different analysis techniques, AMANDA and NT-200
yield limits on the diffuse
flux of neutrinos with extraterrestrial origin for neutrino energies
from 6 TeV up to a few EeV. Figure \ref{fig:diffuse}
summarizes the flux predictions, upper bounds derived
from observed fluxes of charged cosmic rays and gamma rays,
and existing best limits of AMANDA and NT-200 as well as
extrapolations to several years of AMANDA data and to IceCube.
With the exception of the limit from the unfolded atmospheric spectrum,
which can be seen as a quasi-differential limit, the limits are on the
integrated flux over the energy range which contains 90\% of the signal.
These limits exclude some models like \cite{SS}. Note that
both experiments enter new territory, 
being below the gamma bound \cite{gamma}
where sensitivities are not {\it a priori} too weak to hope
for a discovery (see also section 4.2).

\subsection{Search for point sources}

\begin{figure}[hb]
\begin{center}
\includegraphics[width=7.7cm]{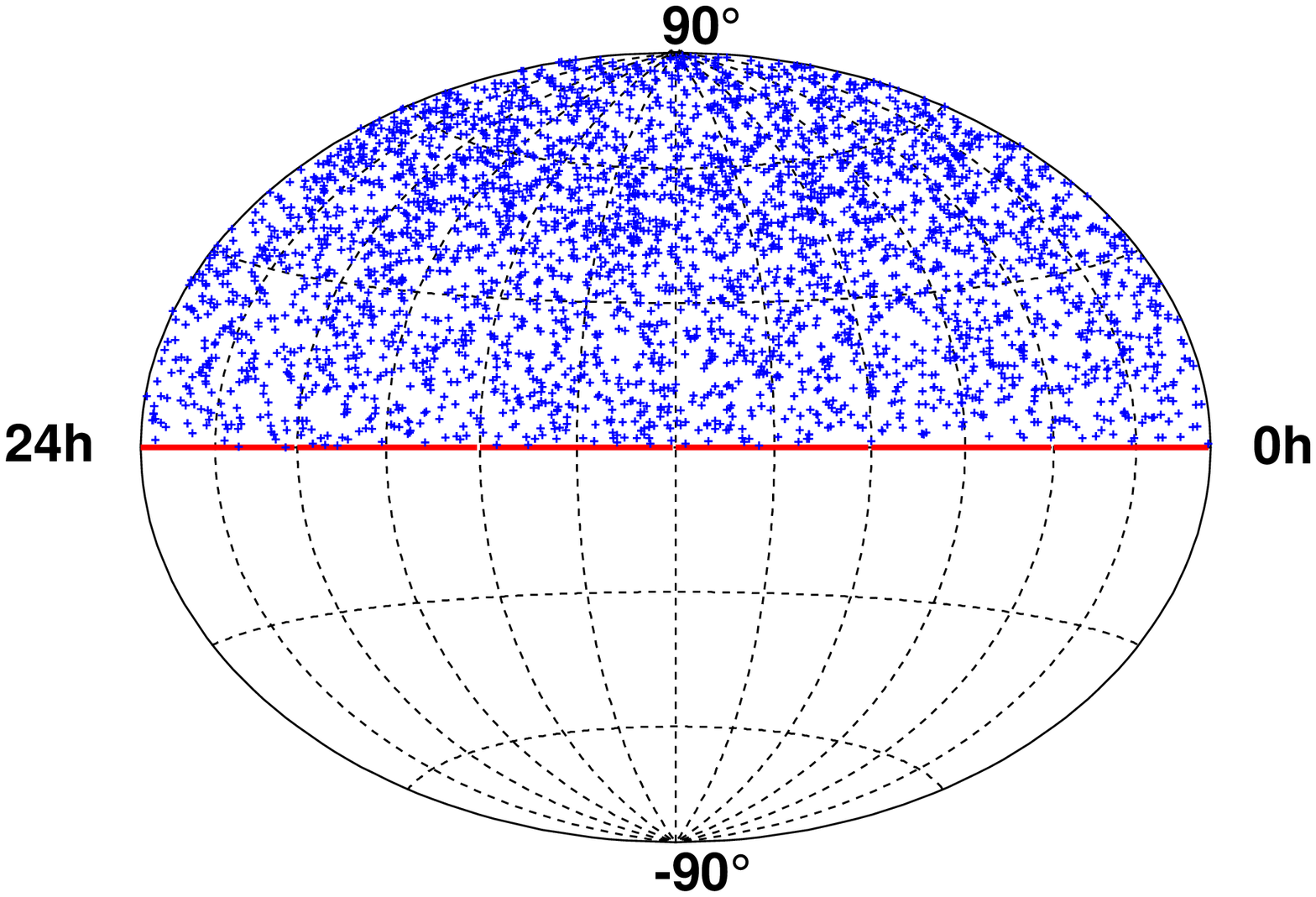}\\
\vspace{-11mm}
\hspace{2mm}
\includegraphics[width=7.7cm,height=7.3cm]{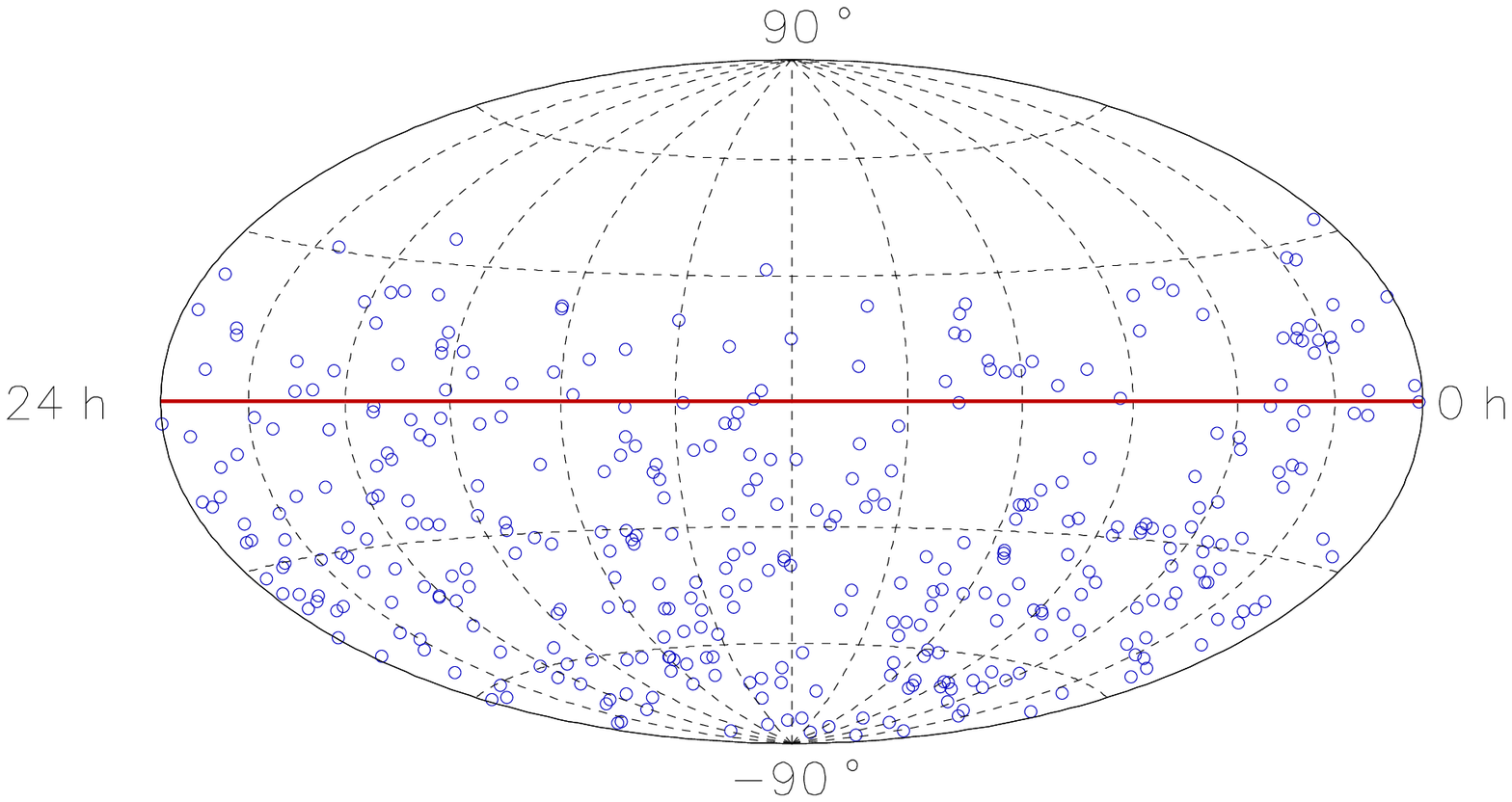}
\vspace{-10mm}
\caption{Sky map in equatorial coordinates. {\it Top:} 3329 neutrino
candidates recorded with AMANDA-II in 2000-2003. {\it Bottom:}
372 neutrino candidates recorded with NT-200 in 1998-2003 (preliminary).
}
\label{fig:skyplots}
\end{center}
\end{figure}

Searches for neutrino 
point sources require good pointing resolution and are
thus restricted to the $\nu_\mu$ channel. Both experiments
have produced sky-plots based on data taken over several years.
Figure \ref{fig:skyplots} shows 3329 upward moving muons recorded
by AMANDA during four years (2000-2003), and 372 Baikal events 
recorded during 5 years (1998-2003).  
Since good pointing resolution is
mandatory, and since this can be achieved only for tracks crossing
the array, the smaller NT-200 cannot compete with AMANDA. We note,
however, that for hard neutrino spectra (e.g. $E^{-2}$) NT-200 is 
yet the largest telescope at the Northern hemisphere and nicely
complements AMANDA in the South. In the following, only AMANDA data
will be discussed in more detail.

\vspace{-2mm}

AMANDA events were selected to maximize the model rejection 
potential for an $E^{-2}$ neutrino spectrum convoluted 
with the background spectra due
to atmospheric neutrinos and misreconstructed atmospheric muons.
The {\it sensitivity} of the analysis, defined as the average upper 
limit one would expect
to set on a non-atmospheric neutrino flux if no signal is detected, 
is about  $E^2 \Phi_{\nu}(E) < 0.6 \times 10^{-7} 
\mbox{cm}^{-2}\mbox{s}^{-1}\mbox{GeV}$, averaged over the lower
hemisphere and
for a hypothetical $E^{-2}$ signal spectrum. This is three
times lower than the one-year limit for the year 2000 
\cite{PS2000}. It corresponds to a neutrino flux above 1 TeV
of $F_{\nu}(>$1~TeV)$ < 6 \cdot 10^{-10}$ cm$^{-2}$~s$^{-1}$ --
about a factor five above the Crab gamma flux and close
to the gamma flux from Markarian-501 in its flaring phase \cite{TeVgamma}.

The final sample of 3369 neutrino candidates (3329 from below, 
with 3438 expected atmospheric neutrinos)
was searched for point sources with two methods.
In the first, the sky is divided into a 
fine-meshed grid of
overlapping bins which are tested for a 
statistically significant excess over the
background expectation (estimated from all other 
bins in the same declination band).
This search yielded no evidence for extraterrestrial point sources.
The second method is an unbinned search, in which the 
sky locations of the events and
their uncertainties from reconstruction are 
used to construct a sky map of significance
in terms of fluctuation over background.
The hot spots on this map are well within
the expectation from a random event distribution. 
In a search for 33 preselected sources,
the strongest excess was observed from the direction of the
Crab nebula, with 10 events where 5 are expected - again
no significant effect given the number of trials.
One thus sees no evidence for point sources 
with an $E^{-2}$ energy spectrum
based on the first four years of AMANDA-II data, and it seems
unlikely that another few years of data will change this
result. Obviously, a much larger array like
IceCube is neccessary to detect
{\it steady} sources. However, the picture may change principally
for {\it transient} sources, where, for searches during known 
gamma flares (like e.g. the blazars
Mkr-421, Mkr-501, ES1959+650), the signal-to-noise ratio may improve 
dramatically.

\subsection{Search for neutrinos from GRBs}

A special case of point source analysis is the search for neutrinos
coincident with gamma ray bursts (GRBs) detected by satellite-borne
detectors. Here, the timing of the neutrino event serves as an additional
selection handle which significantly reduces background.

Both collaborations have used the GRB sample collected by the 
BATSE satellite detector which was decomissioned in 2000.
The AMANDA (Baikal) and BATSE data taking periods were 
overlapping in 1997-2000 (1998-2000).
Samples  containing 312 (368) bursts triggered by BATSE
from this period have been analyzed by AMANDA (Baikal).
Data were searched for an excess 
of events in a 10 min (100 second) window around the GRB time.
No coincident neutrino event was observed in the case of AMANDA,
and one (over a background of 0.46) for Baikal.
Assuming a broken power-law energy spectrum as proposed by Waxmann and Bahcall
\cite{WB}, the 90\% C.L.\ upper limit on the expected neutrino flux 
at the Earth derived by AMANDA is
$E^2 \Phi_{\nu}(E) < 4 \times 10^{-8} 
\mbox{cm}^{-2}\mbox{s}^{-1}\mbox{sr}^{-1}\mbox{GeV}$.
This is approximately a factor 15 above the Waxmann-Bahcall 
flux prediction.

Other classes of bursts are being included in the analysis, like
the so-called non-triggered BATSE bursts and triggers
from the 
Third Interplanetary Network (IPN3), since
2000 the major source of GRB detection.

\subsection{Dark matter search}

The Minimal Supersymmetric extension of the Standard Model
(MSSM) provides a promising dark matter candidate 
in the neutralino, which could be the
lightest supersymmetric particle.
Neutralinos can be gravitationally trapped in massive bodies, and can then via
annihilations and the decay of the resulting particles produce neutrinos.
Dark matter can therefore indirectly searched for by
looking for fluxes of neutrinos from the center of the Earth or the Sun.

\begin{figure}[h]
\begin{center}
\includegraphics[width=7.3cm]{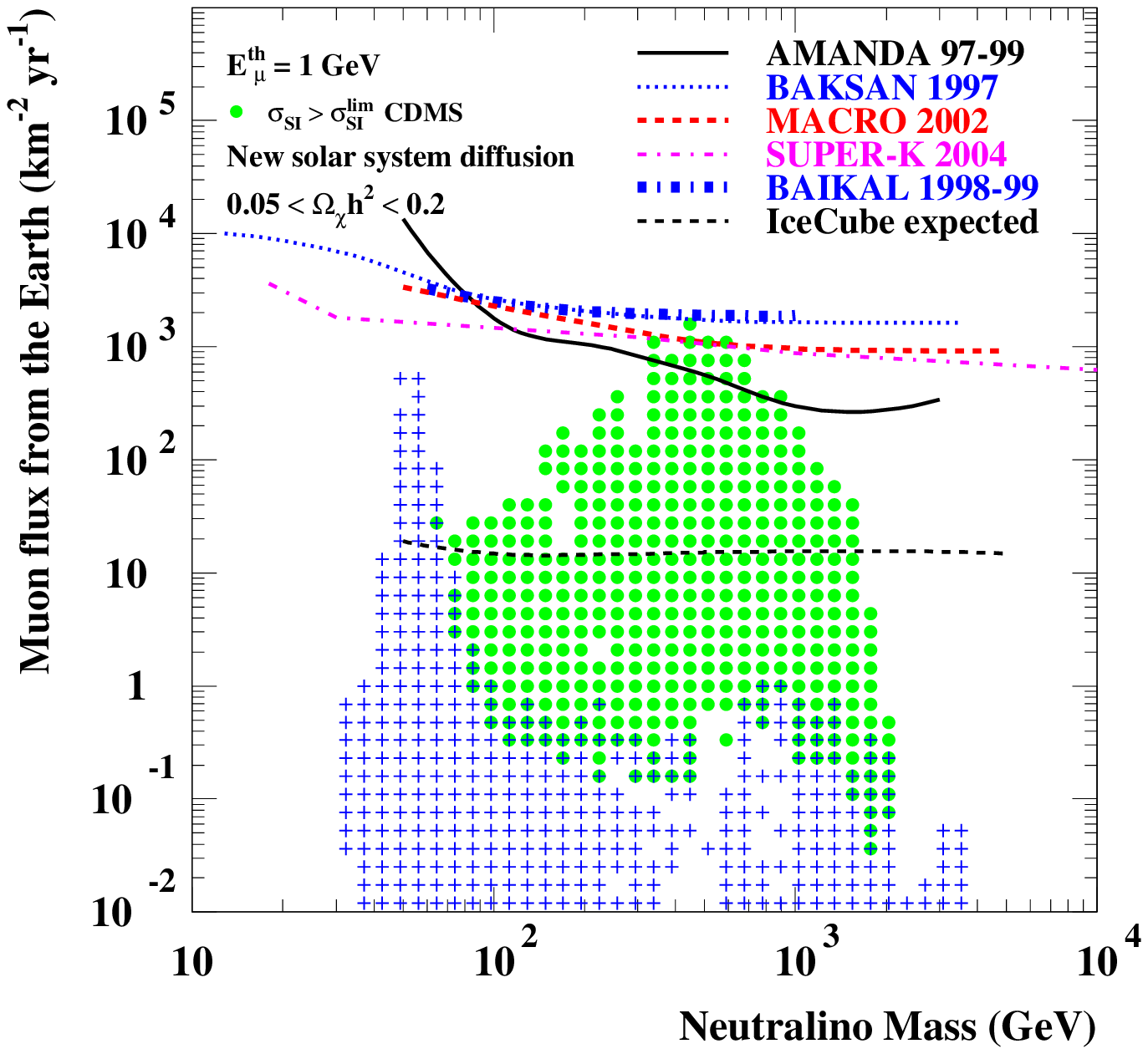}
\hspace{3mm}
\includegraphics[width=7.3cm]{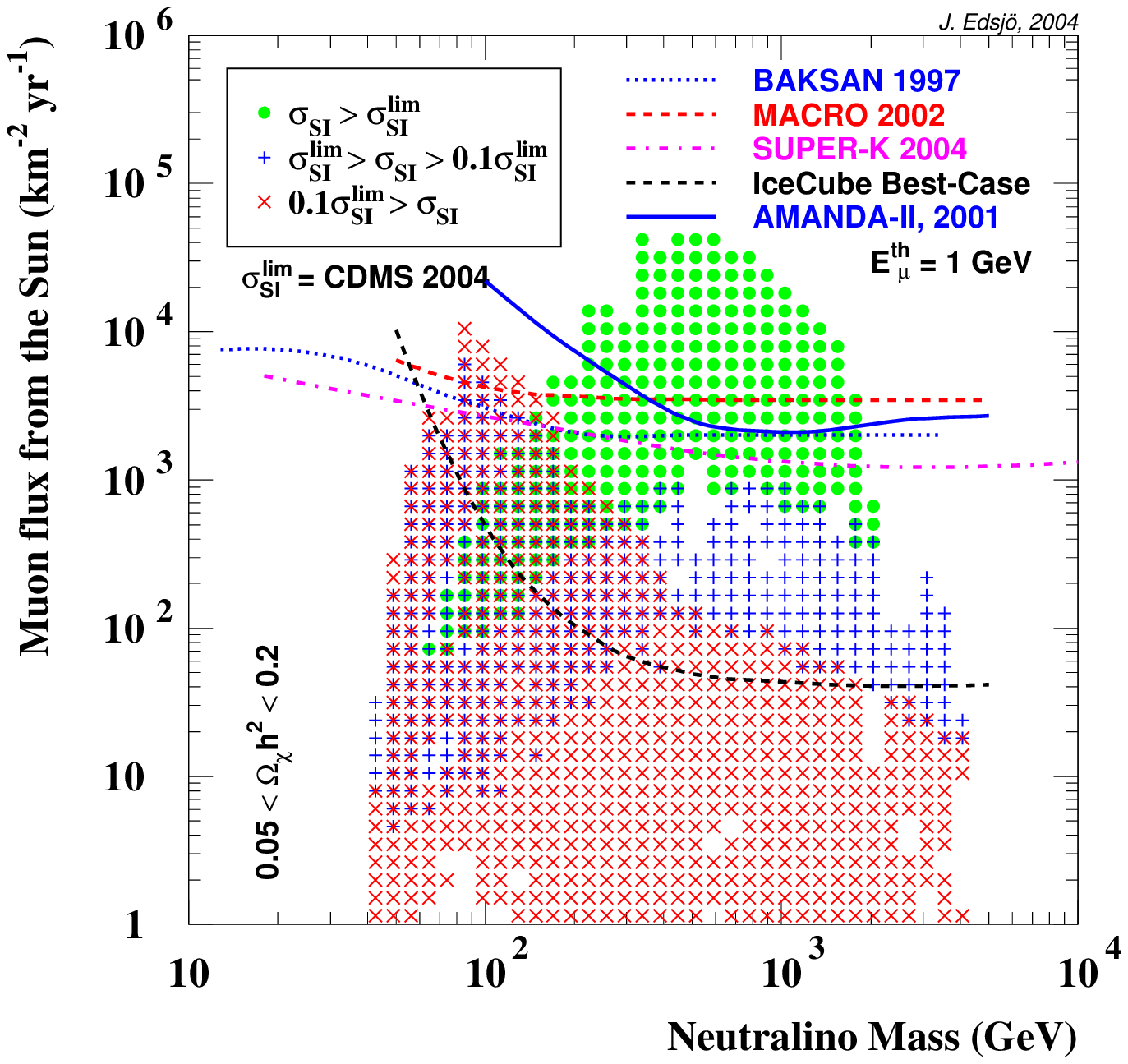}
\caption{Limits on the muon flux due to neutrinos from
neutralino annihilations in the center of the Earth (left) 
and the Sun (right). The
symbols correspond to model predictions 
\cite{WIMP} within the allowed
parameter space of the MSSM.
The dots represent models disfavored by direct searches 
with CDMS II~\cite{CDMS2004}. 
Crosses (+) cover models which could be tested
by the direct searches ten times more sensitive than the
present ones. Amanda and Baikal limits are preliminary.
}
\label{fig:wimplimits}
\end{center}
\end{figure}

Both 
collaborations have searched  for vertically up-going tracks
from the center of the Earth, for
AMANDA-B10 using data from
1997-1999, for NT-200 using data from 1998/1999.
No indication of an excess over atmospheric neutrinos
was found. The 90\% C.L.\ upper limit on the muon flux
from the center of the Earth are compared in
Fig.~\ref{fig:wimplimits}~(left) to limits obtained from underground
experiments and to MSSM model predictions excluded by direct
search.

With its larger mass and
a higher capture rate due to additional spin-dependent processes,
the Sun is more effective than the Earth in catching WIMPs.
AMANDA-II data from 2001 data yield no indication of a
WIMP signature.
The preliminary upper limit on the muon flux from the Sun is compared
to MSSM predictions in 
Fig.~\ref{fig:wimplimits} (right).
For heavier neutralino masses, 
the limit obtained with less than one year of AMANDA-II
data is already competitive with limits from indirect searches with detectors
that have several years of integrated livetime.
It should be noted that the two methods are complementary since they
(a) probe the WIMP distribution in the solar system at different epochs
and (b) are sensitive to different parts of the velocity distribution.
Direct searches are sensitive to high-energy recoils and 
therefore to the high-velocity tail of the WIMP flux, indirect 
searches are more sensitive to low-velocity WIMPs since those
are easier captured by celestial bodies.

\vspace{-2mm}

\subsection{Magnetic monopoles}

\begin{figure}[htb]      
\begin{center}
\includegraphics[width=9.2cm]{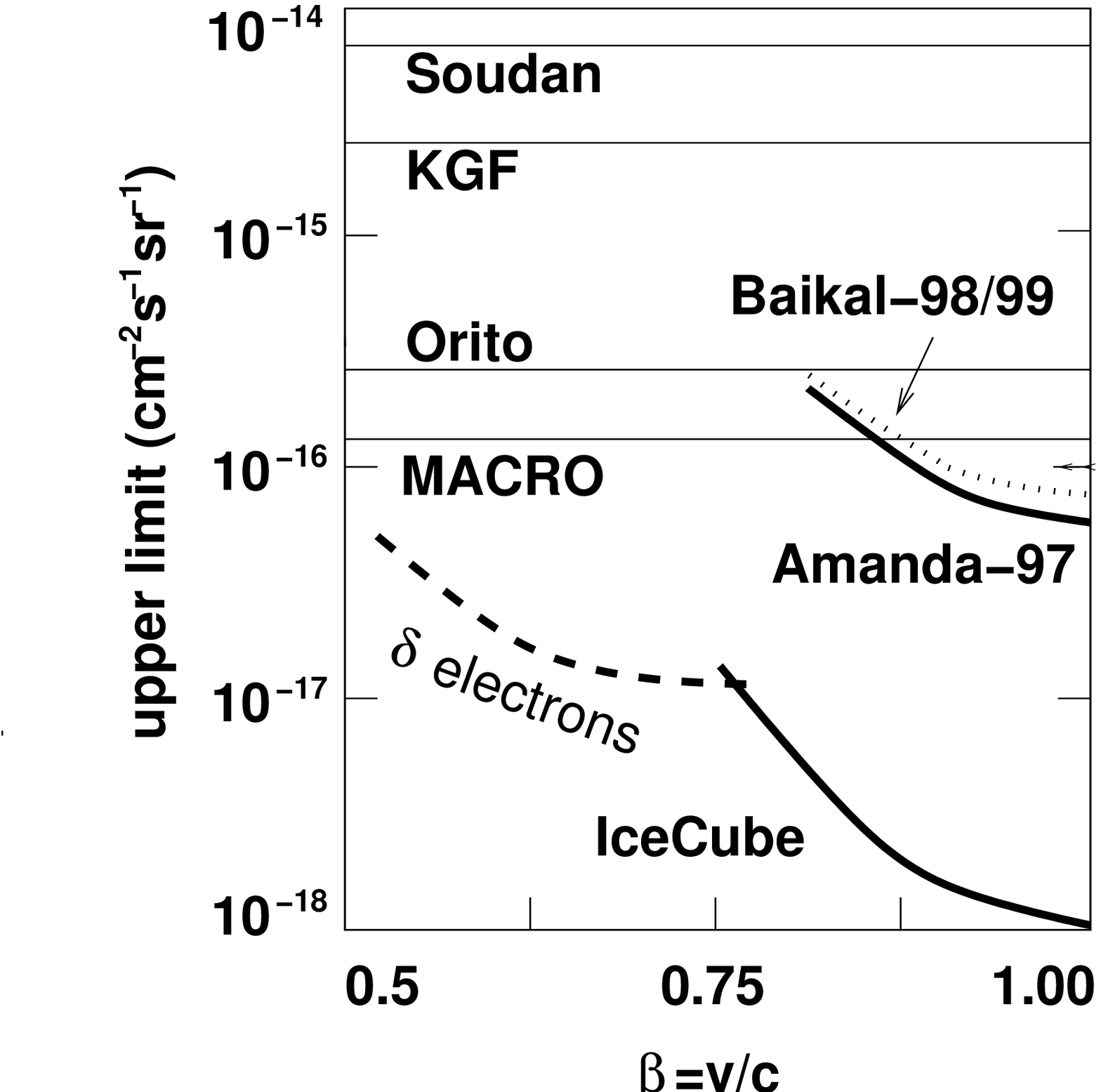}
\caption{Upper limits on the flux of relativistic monopoles
obtained in different experiments.
}
\label{fig:monopol}
\end{center}
\end{figure}

A  magnetic monopole with unit magnetic Dirac charge $g = 137/2 \cdot e$
and velocities above the Cherenkov threshold in water
$(\beta > 0.75)$ would emit 
Cherenkov radiation smoothly along its path, 
exceeding that of a bare relativistic muon 
by a factor of 8300. This is
a rather unique signature. Figure \ref{fig:monopol} summarizes 
the limits obtained until now. A cube kilometer detector could 
improve the sensitivity of this search by nearly two orders of 
magnitude. The search could be extended to even lower velocities 
by detection of the $\delta$ electrons generated along the monopole path.

\bigskip

Limits on the flux of particles moving with with less than
$10^{-3} c$, like GUT magnetic monopoles catalyzing
baryon decay, Q-balls or nuclearites
\cite{MM} have been obtained with early stages of the Baikal
detector \cite{NT-36,Girlanda86}. The trigger system of the future
Icecube is flexible enough to search effectively for such
particles and to lead to much stronger limits
than those of \cite{Girlanda86}.

\subsection{Detection of Supernova bursts}

Due to the low external noise rate, AMANDA is sensitive to
the increase of individual counting rates of all PMs  
resulting from a Supernova burst \cite{B10sn}.
AMANDA-II can detect 90\% of supernovae within
9.4 kpc with less than 15 fakes per year.
This is sufficiently robust for AMANDA to 
contribute to the SuperNova Early Warning System
(SNEWS) with neutrino detectors in the Northern hemisphere.
IceCube will monitor the full Galaxy. An alarm
would confirm other records within SNEWS
but also provide directional information: If several detectors spread 
around the world would measure the signal front with an accuracy of a 
few ms, one might determine the supernova direction by 
triangulation. The high statistics
of hits recorded by IceCube would allow a precise
measurement of the set-in of the burst \cite{Triang}.

\section{IceCube}

\subsection{The detector}

\begin{figure}[ht]      
\begin{center}
\includegraphics[width=8cm,angle=-90]{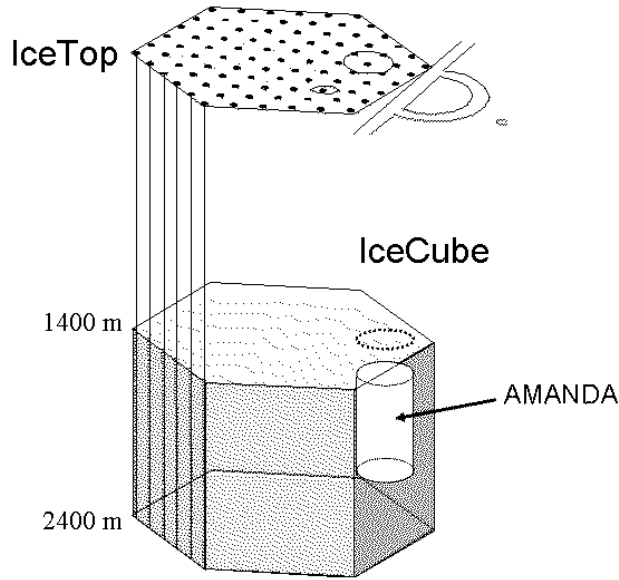}
\caption{Schematical View of IceCube, IceTop and AMANDA.
}
\label{fig:icecube}
\end{center}
\end{figure}

With 4800 optical modules on 80 strings, horizontally
spaced by 125 m, IceCube \cite{IceCube1}
covers an area of approximately 1 km$^2$,
with the OMs at depths of 1.4 to 2.4 km below surface. Each string
carries 60 OMs, vertically spaced by 17 m. The strings are
arranged in a triangular pattern. 
The  configuration of IceCube is shown in Fig.\ref{fig:icecube}.
At each hole, one
station of the IceTop air shower array \cite{Stanev}
will be positioned.
An IceTop station consists of two ice tanks of total area
7 m$^2$.  

A first IceCube string has been successfully deployed
on January 27, 2005. For IceCube,
a new drilling system 
has been constructed. The power for heaters and pumps
of the EHWD is now $\sim$ 5 MW, compared to 2 MW for AMANDA.
This, and the larger diameter and length of the water transporting
hoses, results in only 40 hours needed to drill a
2400 m deep hole (three times faster than with the old
AMANDA drill).  Mounting, testing and drop of a string with 60 DOMs 
takes about 20 hours, and deployment of ultimately 18 strings per season
seems feasible.

The present AMANDA-II detector 
will be integrated into IceCube. 
IceCube will deliver efficient
veto information for low energy cascade-like events 
or short horizontal tracks recorded in AMANDA. Horizontal
tracks could  be related to neutrinos steming from WIMP 
annihilations in the Sun.

The IceCube OM contains 
a 10-inch diameter PM HAMAMATSU R-7081.
Different to AMANDA, 
the PM anode signal 
is digitized within the OM   
and sent to the surface via electrical 
twisted-pair cables. 
Waveforms of the signals are
recorded with 250 MHz over the first 0.5 $\mu$s
and 40 MHz over 5 $\mu$s.
The fine sampling is done with 
the Analog Transient Waveform Recorder (ATWR), an ASIC with four
channels, each capable to capture 128 samples with 200-800 Hz.
The 40 MHz sampling is performed by a commercial FADC.
Each pulse time 
stamped with 7 ns r.m.s.. 
See \cite{IceCube} for more details.

\subsection{Physics Performance}

\begin{figure}[ht]      
\begin{center}
\includegraphics[width=7.5cm]{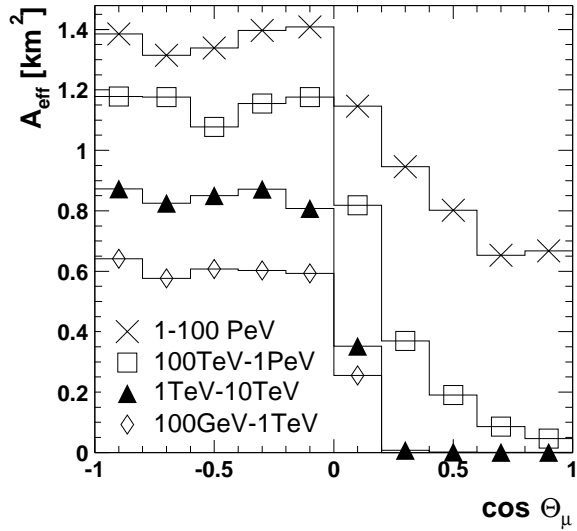}
\caption{IceCube
effective area for muon as a function of zenith angle.
$\cos\theta=-1$ denotes vertically upward moving muons,
$\cos\theta=0$ marks the horizontal direction.
}
\label{fig:area}
\end{center}
\end{figure}

Figure \ref{fig:area} shows the IceCube effective 
area for muons after $\sim 10^{-6}$ reduction 
of events from downward muons,
as a function of the muon zenith angle \cite{IceCubeapp}. Whereas at 
TeV energies IceCube is blind towards the upper hemisphere, 
at PeV and beyond the aperture extends above the horizon and 
allows observation of the Southern sky.

The IceCube sensitivity to diffuse fluxes
after three years of data taking
is shown in Fig.~\ref{fig:diffuse} (section 3.2).
The dashed-dotted lines indicate the 
Stecker and Salamon model for photo-hadronic interactions in AGN 
cores \cite{SS} and of the model of
Mannheim, Protheroe and Rachen 
on neutrino emission from photo-hadronic interactions in AGN 
jets \cite{MPR}. In case of
no signal observed, these models could be rejected with 
model rejection factor of
$10^{-3}$ and $10^{-2}$ respectively. Also shown is the GRB
estimate by Waxman and Bahcall \cite{WB}
which would yield of the order of ten events coinciding with
a GRB, for 1000 monitored GRBs.

\begin{figure}[h]      
\begin{center}
\includegraphics[width=8.5cm]{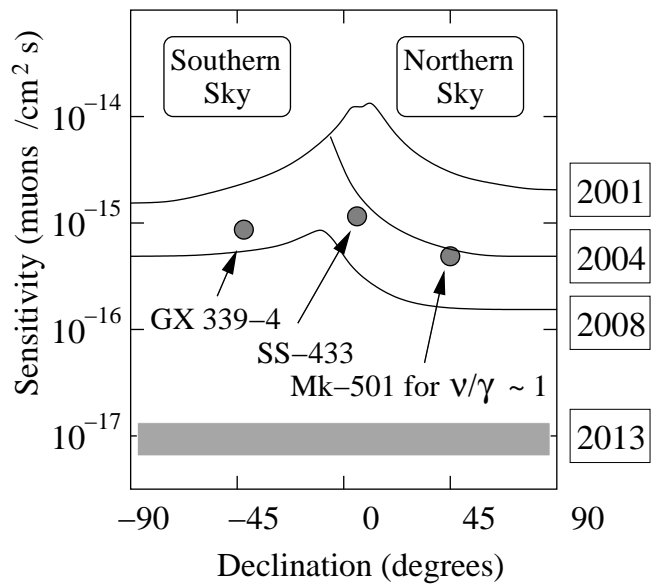}
\caption{
Scenario for the 
improvement of experimental sensitivities to TeV point sources. 
Expected steps for the Northern sky are obtained from Amanda
alone  
and Amanda together with the first 25-30 strings of IceCube, 
on the Southern sky from the Mediterranean detectors Antares
and Nestor (2008). 
In 2013, both hemispheres will be
instrumented with cubic 
kilometer arrays indicated by the grey band.
Shown are also predicted fluxes for two microquasars \cite{Guetta} - one at
northern and one on the southern hemisphere - which are just in
reach for Amanda and the Mediterranean arrays. As a benchmark,
I also show the flux which would be expected if Mk-501, a source
spectacular in TeV gamma rays,  would produce a 
flux of TeV muon neutrinos similar to that of gammas emitted
in its flaring phase.
}
\label{fig:pointfuture}
\end{center}
\end{figure}

For not too steep angles the IceCube pointing
resolution is 0.6 to 0.8 degrees, improving with energy. We expect that
evaluation of waveform information will improve these numbers significantly,
at least at high energies, and increase the potential for
point source identification.  

Figure \ref{fig:pointfuture} 
sketches a possible scenario for the point source search over
the next decade.
Best present limits are from MACRO, Super-Kamiokande (Southern sky) 
and AMANDA (Northern sky). Baikal limits for
the Southern sky will appear soon. 
This picture will not change until 
the medium stage Mediterranean detectors come into operation.
The ultimate sensitivity for 
the TeV-PeV range is likely reached by the cubic kilometer arrays. 
This scale is set by many model predictions for neutrinos from cosmic 
accelerators 
or from dark matter decay. 
A discovery with AMANDA is not yet excluded --
be it a flaring blazar or a supernova; for 
IceCube, hundred times larger than AMANDA and thousand 
times larger than underground detectors,
it seems extremely likely.
However, irrespective of 
any specific model  prediction, IceCube 
will hopefully also keep the promise for  
any detector opening  a new window to the Universe: 
to detect {\it unexpected} phenomena.

\section*{Acknowledments} 

I thank my colleagues in the 
collaborations Baikal, AMANDA and IceCube for the long,
fruitful cooperation and for helpful discussions.

\end{document}